\documentclass[aps,pra,twocolumn,groupaddress]{revtex4}
\usepackage{graphicx}
\usepackage{amsmath,amssymb}
\usepackage{bm}
\usepackage{subfigure}
\usepackage{color}
\usepackage{bbold}
\begin{document}
\title{Phase fluctuations and first-order correlation functions of dissipative Bose-Einstein condensates}
\author{A.-W. de Leeuw}
\email{A.deLeeuw1@uu.nl}
\affiliation{Institute for Theoretical Physics and Center for Extreme Matter and Emergent Phenomena, Utrecht
University, Leuvenlaan 4, 3584 CE Utrecht, The Netherlands}
\author{H.T.C. Stoof}
\affiliation{Institute for Theoretical Physics and Center for Extreme Matter and Emergent Phenomena, Utrecht
University, Leuvenlaan 4, 3584 CE Utrecht, The Netherlands}
\author{R.A. Duine}
\affiliation{Institute for Theoretical Physics and Center for Extreme Matter and Emergent Phenomena, Utrecht
University, Leuvenlaan 4, 3584 CE Utrecht, The Netherlands}
\date{\today}
\begin{abstract}
We investigate the finite lifetime effects on first-order correlation functions of dissipative Bose-Einstein condensates. By taking into account the phase fluctuations up to all orders, we show that the finite lifetime effects are neglible for the spatial first-order correlation functions, but have an important effect on the temporal correlations. As an application, we calculate the one-particle density matrix of a quasi-condensate of photons. Finally, we also consider the photons in the normal state and we demonstrate that the finite lifetime effects decrease both the spatial and temporal first-order correlation functions. 
\end{abstract}
\maketitle
\vskip2pc
\section{Introduction}
\label{sec:int}
Bose-Einstein condensation (BEC) was first directly observed by cooling atomic vapors to temperatures in the nK regime \cite{BEC1,BEC2,BEC3}. These vapors were confined in three-dimensional traps, of which the shape can be altered by changing certain experimental parameters, e.g. the magnetic field. For sufficiently tight confinement in two or one directions, the dynamics in these dimensions can be frozen out and the atoms behave as a quasi one-dimensional or two-dimensional gas. This manipulation opened up the possibility to explore Bose-Einstein condensates in lower dimensions \cite{Gor,Schreck,Ott,Hansel}.
\newline
\indent From a theoretical point of view these low-dimensional Bose-Einstein condensates are particulary interesting as their physics is fundamentally different from three-dimensional condensates. Namely, in two dimensions a homogeneous Bose gas can only undergo BEC at zero temperature and in one dimension BEC in a homogeneous Bose gas cannot take place at all \cite{Mermin,Hohenberg}. In the presence of an external potential the situation drastically changes. In particular, harmonically trapped bosons can undergo BEC at non-zero temperatures in both one and two dimensions \cite{Ketterle,Bagnato,BECharm2}. 
\newline
\indent For homogeneous two-dimensional Bose gases, theoretical studies show that even though the Bose-Einstein condensate does not exist at non-zero temperatures, there still exists a critical temperature in the system. Below the so-called Kosterlitz-Thouless temperature the gas is superfluid, and above this temperature the bosons lose their superfluid property \cite{Kost}. This is known as the Kosterlitz-Thouless transition, and it implies that superfluidity only requires the presence of a quasi-condensate \cite{Popov} with phase coherence over a distance much less than the system size. This quasi-condensate can be roughly interpreted as a system consisting of several patches with each a fixed phase, whereas the phases of these different patches are uncorrelated.  
\newline
\indent In addition to atomic gases, there are presently also other low-dimensional systems in which BEC is observed, such as systems consisting of exciton-polaritons \cite{BECpolariton,BECpolariton2} or photons \cite{BECphoton}. Together with BEC of magnons \cite{BECmagnon}, these systems form a class of condensates that is different from the  atomic Bose-Einstein condensates. In particular, the bosonic quasiparticles have a small effective mass resulting in BEC at temperatures in the range of 10 - 300 K instead of in the nK regime relevant for the atomic Bose-Einstein condensates. Furthermore, these condensates are not in true thermal equilibrium, and the steady state is a dynamical balance between particle losses and external pumping. Therefore, the particles have a finite lifetime, which can be characterized by a single dimensionless damping parameter. For BEC of photons this is explicitly shown in Ref.\,\cite{AW}. 
\newline
\indent In this article we study the effect of this damping parameter on the first-order correlation functions of low-dimensional Bose-Einstein condensates. First, we derive a general expression for the first-order correlation function for a homogeneous Bose gas in the condensed phase in Sec.\,\ref{sec:phasefluc}. Hereafter, we use this general expression for the first-order correlation functions to determine the effect of the finite lifetime on the spatial and temporal correlations in Sec.\,\ref{sec:nonequi}. In Sec.\,\ref{sec:normal} we focus on BEC of photons, taking their interaction with the dye molecules into account, and determine the off-diagonal long-range behaviour of the one-particle density matrix in the Bose-Einstein condensed phase. We show that for the relevant parameters used in the experiment the photons form a true condensate. Subsequently, we determine the first-order correlation functions of photons in the normal state, and we end with conclusions and an outlook in Sec.\,\ref{sec:concl}.  

\section{Phase fluctuations}
\label{sec:phasefluc}
In this section we derive a general expression for the first-order correlation functions for a homogeneous Bose gas consisting of $N$ bosons in a box of volume $V$. We start from the Euclidean action
\begin{align}\label{eq:act}
S&[\phi^{*}, \phi] = \int_{0}^{\hbar\beta} d\tau \, d\tau^{\prime} \int d{\bf x} \, d{\bf x}^{\prime} \, \phi^{*}({\bf x}, \tau) \\ \nonumber
&\times \Bigg{\{} \left[\hbar \frac{\partial}{\partial \tau} - \frac{\hbar^{2}}{2 m} \nabla^{2} - \mu + \frac{1}{2} T^{\mathrm{2B}} |\phi({\bf x}, \tau)|^{2}\right] \delta(\tau - \tau^{\prime}) \\ \nonumber
&\times \delta({\bf x} - {\bf x}^{\prime}) + \hbar\Sigma({\bf x} - {\bf x}^{\prime},\tau - \tau^{\prime}) \Bigg{\}} \phi({\bf x}^{\prime}, \tau^{\prime}) ,
\end{align}
where $\tau$ and $\tau^{\prime}$ denote imaginary times, $\mu$ is the chemical potential and $T^{\mathrm{2B}}$ is the strength of the self-interaction. Furthermore, we included a self-energy $\hbar\Sigma({\bf x} - {\bf x}^{\prime},\tau - \tau^{\prime})$ describing additional interaction effects, e.g. in the photon experiment of Klaers \emph{et al.}\ the interaction of the photons with the dye molecules \cite{BECphoton}. 
\newline
\indent In order to obtain an expression for the first-order correlation functions in the superfluid phase, we split the density and phase fluctuations and substitute $\phi({\bf x}, \tau) = \sqrt{n + \delta n({\bf x}, \tau)} e^{i \theta({\bf x}, \tau)}$. Here $n$ is the average density of the gas, $\delta n({\bf x}, \tau)$ denotes the density fluctuations and $\theta({\bf x}, \tau)$ represents the phase. We expand up to second order in $\theta$ and $\delta n$, and define
\begin{align}\label{eq:FT}
\theta({\bf x},\tau) &= \frac{1}{\sqrt{\hbar\beta V}} \sum_{{\bf k},m} \theta_{{\bf k},m} e^{i ({\bf k} \cdot {\bf x} - \omega_{m} \tau)},\\ \nonumber
\delta n({\bf x},\tau) &= \frac{1}{\sqrt{\hbar\beta V}} \sum_{{\bf k},m} \delta n_{{\bf k},m} e^{i ({\bf k} \cdot {\bf x} - \omega_{m} \tau)},
\end{align}
to obtain 
\begin{align}\label{eq:action5}
S&[\delta n, \theta] = \frac{n}{2} \sum_{{\bf k},m} \left\{\hbar\Sigma^{\mathrm{s}}({\bf k},i \omega_{m}) + 2 \epsilon({\bf k}) \right\} \theta_{{\bf k},m} \theta_{-{\bf k},-m} \\ \nonumber
&+ \frac{1}{8n} \sum_{{\bf k},m} \left\{\hbar\Sigma^{\mathrm{s}}({\bf k},i \omega_{m}) + 4 n \chi^{-1}({\bf k}) \right\} \delta n_{{\bf k},m} \delta n_{-{\bf k},-m} \\ \nonumber
&+\sum_{{\bf k},m} \left\{\hbar\omega_{m} + \frac{i}{2}\hbar\Sigma^{\mathrm{a}}({\bf k},i \omega_{m}) \right\} \delta n_{-{\bf k},-m} \theta_{{\bf k},m} 
\end{align}
with the inverse of the static density-density correlation function $\chi^{-1}({\bf k}) = \epsilon({\bf k})/2n + T^{\mathrm{2B}}$, the single-particle dispersion $\epsilon({\bf k}) = \hbar^{2} {\bf k}^{2}/2 m$ and the antisymmetric and symmetric parts of the self-energy obeying
\begin{align}
\hbar\Sigma^{\mathrm{a}}({\bf k},i \omega_{m}) &= \hbar\Sigma({\bf k},i \omega_{m}) - \hbar\Sigma(-{\bf k},-i \omega_{m}), \\ \nonumber
\hbar\Sigma^{\mathrm{s}}({\bf k},i \omega_{m}) &= \hbar\Sigma({\bf k},i \omega_{m}) + \hbar\Sigma(-{\bf k},-i \omega_{m}).
\end{align} 
Here $\hbar\Sigma({\bf k},i \omega_{m})$ is defined in a similar way as the Fourier transform of the phase and density fluctuations in Eq.\,\eqref{eq:FT} expect for the normalization factor which is $1/ \hbar \beta V$ in this case. Furthermore, in Fourier space we take without loss of generality $\hbar\Sigma({\bf 0},0) = 0$. For bosons this assumption is automatically satisfied for the imaginary part, and the constant real part results in a energy shift of the poles of the Green's function and can be absorped in the chemicial potential. Note that in Eq.\,\eqref{eq:action5} we substituted the zero-loop result for the chemical potential $\mu = n T^{\mathrm{2B}}$. 
\newline
\indent By using the classical equations of motion, we can now eliminate the phase $\theta_{{\bf k},m}$ and find an action for the density fluctuations $\delta n_{{\bf k},m}$ alone. From this action we obtain
\begin{align}\label{eq:cordens1}
\langle &\delta n({\bf x}, \tau) \delta n({\bf x}^{\prime}, \tau^{\prime}) \rangle \\ \nonumber 
&=\frac{n}{\beta V} \sum_{{\bf k},m} \frac{\hbar\Sigma^{\mathrm{s}}({\bf k}, i \omega_{m}) + 2 \epsilon({\bf k})}{ \mathrm{Det}[\hbar {\bf G}_{\mathrm{B}}^{-1}({\bf k},i \omega_{m})] } e^{i ({\bf k} \cdot ({\bf x} - {\bf x}^{\prime}) - \omega_{m} (\tau - \tau^{\prime}))},
\end{align}
where the Green's function of the density and phase fluctuations has the matrix structure
\begin{align}\label{eq:fluc}
&\hbar {\bf G}^{-1}_{\mathrm{B}}({\bf k}, i\omega_{m}) \\ \nonumber
&=
\left [ \begin{array} {cc}
\hbar G^{-1}({\bf k}, i\omega_{m}) - n T^{\mathrm{2B}} & - n T^{\mathrm{2B}} \\
- n T^{\mathrm{2B}} & \hbar G^{-1}(-{\bf k}, -i\omega_{m}) -n T^{\mathrm{2B}}
\end{array} \right ],
\end{align}
in terms of the single-particle Green's function
\begin{align}\label{eq:grbog}
\hbar G^{-1}({\bf k}, i\omega_{m}) = i \hbar \omega_{m} - \epsilon({\bf k}) - \hbar\Sigma({\bf k}, i \omega_{m}).
\end{align}
\indent Similarly we use the equation of motion for $\delta n_{{\bf k},m}$ to eliminate the density fluctuations, and we find
\begin{align}\label{eq:corphase1}
\langle &\theta({\bf x}, \tau) \theta({\bf x}^{\prime}, \tau^{\prime}) \rangle = \frac{1}{4 n \beta V} \\ \nonumber
& \sum_{{\bf k},m} \frac{\hbar\Sigma^{\mathrm{s}}({\bf k},i \omega_{m}) + 2 \epsilon({\bf k}) + 4 n T^{\mathrm{2B}}}{ \mathrm{Det}[\hbar {\bf G}_{\mathrm{B}}^{-1}({\bf k},i \omega_{m})] }e^{i ({\bf k} \cdot ({\bf x} - {\bf x}^{\prime}) - \omega_{m} (\tau - \tau^{\prime}))}.
\end{align}
However, from Ref.\,\cite{Chapter15} we know that the contribution of the phase fluctuations is proportonial to the density. The first two terms are not proportional to $n$ and they are an artifact of making an expansion up to second order in $\theta({\bf x},\tau)$, and neglecting the interaction terms between the density and phase fluctuations. 
\begin{figure*}[t]
  \begin{center}
\setlength{\abovecaptionskip}{17.5 pt plus 4pt minus 2pt}
    \mbox{
      \subfigure{\includegraphics[scale=1.14]{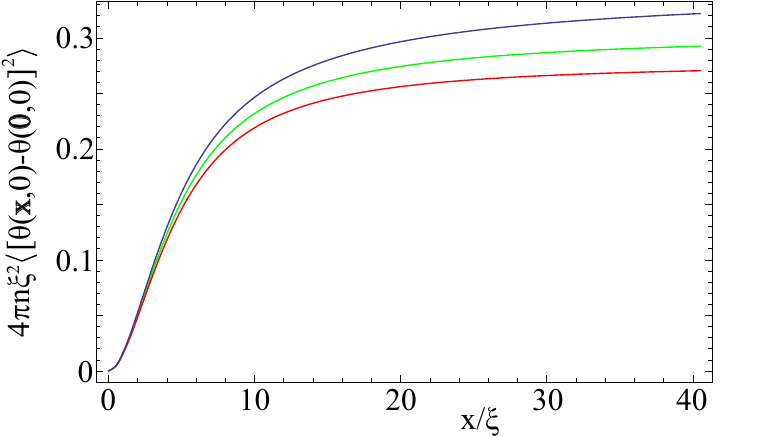}}  \,
      \subfigure{\includegraphics[scale=1.14]{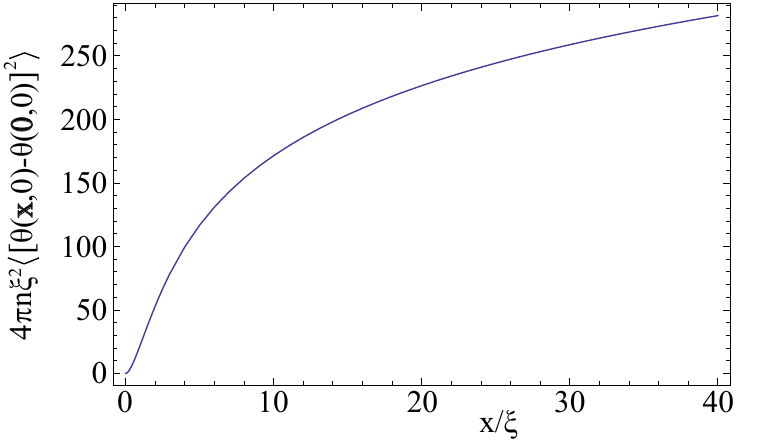}}
      }
\vspace{-0.7 cm}
    \caption{The zero temperature part (left) and the non-zero temperature part (right) of the phase fluctuations in two dimensions for a condensate density $n \simeq 5 \cdot 10^{13} \mathrm{m}^{-2}$ and correlation length $\xi \simeq 2.8 \cdot 10^{-6}\,\mathrm{m}$. In the non-zero temperature part we take $\beta n_{0} T^{\mathrm{2B}} \simeq 1.3 \cdot 10^{-2}$.  The blue, green and red curve correspond to respectively $\alpha = 0$, $\alpha = 5 \cdot 10^{-2}$ and $\alpha = 10^{-1}$.}
     \label{fig:PhaseCorTemp}
  \end{center}
\end{figure*}
A more accurate approach that takes into account higher-order terms in $\theta({\bf x},\tau)$ would not contain these high-momentum contributions. Therefore, the correct expression for the phase fluctuations is given by
\begin{align}\label{eq:corphase2}
\langle &\theta({\bf x}, \tau) \theta({\bf x}^{\prime}, \tau^{\prime}) \rangle \\ \nonumber
&=\frac{1}{n \beta V} \sum_{{\bf k},m} \frac{n T^{\mathrm{2B}}}{ \mathrm{Det}[\hbar {\bf G}_{\mathrm{B}}^{-1}({\bf k},i \omega_{m})] }e^{i ({\bf k} \cdot ({\bf x} - {\bf x}^{\prime}) - \omega_{m} (\tau - \tau^{\prime}))}.
\end{align}
\indent In certain cases the self-energy is only known for real frequencies. Therefore, we define
\begin{align}
\rho_{\theta}({\bf k}, \omega) &=  -\frac{n T^{\mathrm{2B}}}{2 \hbar} \rho_{\mathrm{B}}({\bf k}, \omega)  \\ \nonumber
&:= -\frac{n T^{\mathrm{2B}}}{\pi\hbar} \mathrm{Im}\left[\frac{1}{ \mathrm{Det}[\hbar {\bf G}_{\mathrm{B}}^{-1}({\bf k},\omega^{+})] } \right],
\end{align}
and we write for the phase correlation function
\begin{align}\label{eq:phasegamma7}
\langle &\theta({\bf x}, \tau) \theta({\bf x}^{\prime}, \tau^{\prime}) \rangle \\ \nonumber
&=  \frac{\hbar}{n \hbar \beta V} \sum_{{\bf k},m} \int_{-\infty}^{\infty} d(\hbar\omega) \, \frac{\rho_{\theta}({\bf k}, \omega)}{i \omega_{m} - \omega} e^{-i \omega_{m} (\tau - \tau^{\prime})} e^{i {\bf k} ({\bf x} - {\bf x}^{\prime})} \\ \nonumber
&= -\frac{\hbar}{n V} \sum_{{\bf k}} \int_{-\infty}^{\infty} d(\hbar \omega) \, \rho_{\theta}({\bf k}, \omega) e^{-\omega (\tau - \tau^{\prime})} e^{i {\bf k} ({\bf x} - {\bf x}^{\prime})} \\ \nonumber 
&\times \left\{\Theta(\tau^{\prime} - \tau) N_{{\mathrm{BE}}}(\hbar \omega) + \Theta(\tau - \tau^{\prime}) (N_{{\mathrm{BE}}}(\hbar \omega) + 1) \right\},
\end{align}
where 
\begin{align}\label{eq:BED}
N_{{\mathrm{BE}}}(\hbar\omega) = \frac{1}{e^{\beta \hbar \omega} - 1},
\end{align}
is the Bose-Einstein distribution function. For simplicity we take $\tau^{\prime} > \tau$, and the case $\tau^{\prime} < \tau$ is treated analogously. In principle we have to consider both the density and phase fluctuations, in order to calculate the first-order correlation functions. However, here we consider relatively high condensate fractions for which the density fluctuations are strongly suppressed and the phase flucutations are most important, especially for the description of the long-range order which is of most interest to us here \cite{dens1,dens2,dens3,dens4}. Hence, we have
\begin{align}\label{eq:phasegone}
\langle \phi^{*}({\bf x}, t) \phi({\bf x}^{\prime}, t^{\prime}) \rangle &\simeq n_{0} \langle e^{-i (\theta({\bf x},t) - \theta({\bf x}^{\prime},t^{\prime}))} \rangle \\ \nonumber
&= n_{0} e^{-\frac{1}{2} \langle[\theta({\bf x},t) - \theta({\bf x}^{\prime},t^{\prime})]^{2} \rangle},
\end{align}
with $n_{0}$ the quasi-condensate density. By using Eq.\,\eqref{eq:phasegamma7} and performing the analytical continuation to real time $\tau = i t$, we obtain for the exponent of this expression
\begin{align}\label{eq:phasegamma5}
\langle[&\theta({\bf x},t) - \theta({\bf x}^{\prime},t^{\prime})]^{2} \rangle \\ \nonumber
&= -\frac{T^{\mathrm{2B}}}{V} \sum_{{\bf k}} \int_{-\infty}^{\infty} d(\hbar \omega) \rho_{\mathrm{B}}({\bf k}, \omega) N_{{\mathrm{BE}}}(\hbar \omega)  \\ \nonumber
&\qquad \quad \,\,\, \times \left\{1 -  \cos({\bf k} \cdot ({\bf x} - {\bf x}^{\prime})) \cos(\omega (t - t^{\prime}))  \right\},
\end{align}
where we used that $\hbar\Sigma({\bf k},\omega) = \hbar\Sigma(-{\bf k},\omega)$ for an isotropic system. 
\newline
\indent In order to make further progress, we perform a long-wavelength approximation to the self-energy. As mentioned before, the real part of the self-energy can effectively be absorped in the chemical potential and therefore we neglect this part. The imaginary part is zero for ${\bf k} =  {\bf 0}$ and $\omega = 0$, and therefore for small frequencies the imaginary part is linear in $\omega$. Since the result of Eq.\,\eqref{eq:phasegamma5} is dominated by the contributions for small frequencies, the large frequency behaviour is not visible in the final result. Therefore we can safely assume that the self-energy obeys $\hbar\Sigma({\bf k}, \omega) = \hbar\Sigma^{*}({\bf k}, -\omega)$ for the imaginary part of both the retarded and advanced self-energy. This allows us to rewrite
\begin{align}\label{eq:phasegamma2}
\langle[\theta&({\bf x},t) - \theta({\bf x}^{\prime},t^{\prime})]^{2} \rangle \\ \nonumber
&= -\frac{T^{\mathrm{2B}}}{V} \sum_{{\bf k}} \int_{0}^{\infty} d(\hbar \omega) \rho_{\mathrm{B}}({\bf k}, \omega) \left\{1 + 2 N_{{\mathrm{BE}}}(\hbar \omega) \right\} \\ \nonumber
&\qquad \quad \,\,\, \times \left\{1 -  \cos({\bf k} \cdot ({\bf x} - {\bf x}^{\prime})) \cos(\omega (t - t^{\prime}))  \right\}.
\end{align}
This expression for the phase fluctuations contains an ultraviolet divergence. This divergence is a consequence of not taking into account the proper energy-dependence of the self-interaction of the bosons. For atoms this problem was already encountered in Ref.\,\cite{Chapter15}, and in this case the divergence is handled by appropriate renormalization of the interactions. In our case the form of the divergence is the same, since the self-energy must vanish for large momenta. However, we do not know the exact energy-dependence of $T^{\mathrm{2B}}$. Therefore, the cancellation of the ultra-violet divergence requires us to introduce another energy scale $\gamma n T^{\mathrm{2B}}$ that models the correct energy-dependence of the self-interaction of the bosons. We come back to the precise determination of $\gamma$ in the next section. We thus write,
\begin{align}\label{eq:phasegamma2}
\langle[&\theta({\bf x},t) - \theta({\bf x}^{\prime},t^{\prime})]^{2} \rangle = -\frac{T^{\mathrm{2B}}}{V} \sum_{{\bf k}} \int_{0}^{\infty} d(\hbar \omega) \\ \nonumber
&\times \rho_{\mathrm{B}}({\bf k}, \omega) \left\{1 + 2 N_{{\mathrm{BE}}}(\hbar \omega) - \frac{\hbar \omega_{{\mathrm{B}}}({\bf k})}{\epsilon({\bf k}) + \gamma n T^{\mathrm{2B}}} \right\} \\ \nonumber
&\times \left\{1 -  \cos({\bf k} \cdot ({\bf x} - {\bf x}^{\prime})) \cos(\omega (t - t^{\prime}))  \right\},
\end{align}
with $\hbar \omega_{{\mathrm{B}}}({\bf k}) = \sqrt{\epsilon({\bf k})(\epsilon({\bf k}) + 2 n_{0} T^{\mathrm{2B}})}$ the Bogoliubov dispersion. The integrant in Eq.\,\eqref{eq:phasegamma2} must be positive for all ${\bf k}$ and all temperatures, as this term correspond to the expectation value of $|\theta_{{\bf k},n}|^{2}$. Therefore, we have the restriction that $\gamma \geq 1$. Finally, note that this result is consistent with the expressions found in Ref.\,\cite{Chapter15}.

\section{Correlation functions in the condensed phase}
\label{sec:nonequi}
In the previous section we found an expression for the first-order correlation function by taking into account the phase fluctuations up to all orders. As mentioned before, the phase fluctuations are dominated by the small-frequency contributions and for bosons the imaginary part of the self-energy is linear in $\omega$ for small frequencies. Therefore, in this section we take the retarded self-energy equal to $\hbar\Sigma^{+}({\bf k},\omega) = - i \alpha \hbar \omega$ and we investigate the effect of $\alpha$ on the first-order correlation functions. 

\subsection{Spatial correlations}
From Eq.\,\eqref{eq:phasegamma2} we obtain that the phase fluctuations contain a zero-temperature part, and a contribution that is temperature dependent. For the equal-time phase fluctuations at zero temperature, we have
\begin{align}
\langle[\theta&({\bf x},0) - \theta({\bf 0},0)]^{2} \rangle = -\frac{T^{\mathrm{2B}}}{V} \sum_{{\bf k}} \int_{0}^{\infty} d(\hbar \omega) \\ \nonumber
&\rho_{\mathrm{B}}({\bf k}, \omega) \left\{1 - \frac{\hbar \omega_{{\mathrm{B}}}({\bf k})}{\epsilon({\bf k}) + \gamma n T^{\mathrm{2B}}} \right\} \left\{1 -  \cos({\bf k} \cdot {\bf x}) \right\}.
\end{align}
Without loss of generality we have set ${\bf x}^{\prime} = 0$ and we have put $t$ equal to zero. Before we consider the effect of the self-energy on the spatial correlations, we investigate the effect of $\gamma$. We first consider the case without a self-energy. By writing the sum over ${\bf k}$ as an integral, we find in two dimensions
\begin{align}
\langle[&\theta({\bf x},0) - \theta({\bf 0},0)]^{2} \rangle = \\ \nonumber
&\int_{0}^{\infty} dk \, \frac{1 -  J_{0}(k x)}{4 \pi n \xi^{2}} \left\{\frac{1}{\sqrt{k^{2}+1}} - \frac{2 k}{2 k^{2} + \gamma} \right\} ,
\end{align}
where $J_{0}(k x)$ is the Bessel function of the first kind, $\xi = \hbar / [4 m n_{0} T^{\mathrm{2B}}]^{1/2}$ is the correlation length and $x = |{\bf x}|$. In the limit $x \rightarrow \infty$, the Bessel function vanishes and we obtain
\begin{align}
&\langle[\theta({\bf x},0) - \theta({\bf 0},0)]^{2} \rangle \rightarrow \frac{\log(2 \gamma)}{2 \pi} \frac{m T^{\mathrm{2B}}}{\hbar^{2}}.
\end{align}
Thus, the condensate density is given by the right-hand side of the following equation
\begin{align}
\langle \phi^{*}({\bf x},0) \phi({\bf 0},0) \rangle \rightarrow n_{0} \exp\left\{ -\frac{\log(2 \gamma)}{4 \pi} \frac{m T^{\mathrm{2B}}}{\hbar^{2}} \right\},
\end{align}
where we again considered the limit ${\bf x} \rightarrow \infty$. Therefore, by increasing $\gamma$ we effectively increase the interaction strength and, thereby decrease the condensate density of the gas. This dependence can in principle be used to determine the value of $\gamma$ from experiment. In order to determine the effect of the self-energy we here just fix $\gamma$ and set it equal to $1$. In Fig.\,\ref{fig:PhaseCorTemp} we show the result for the zero-temperature part of the phase fluctuations for different values of $\alpha$. If $\alpha$ increases, the contribution of the phase fluctuations decreases. Therefore, for increasing $\alpha$ we obtain that the quantum depletion of the condensate decreases. 
\newline
\indent For systems at low temperatures this is the dominating contribution. However, here we are interested in a BEC at higher temperatures such as BEC of exciton-polaritons and photons. For these condensates the temperature-dependent part is the most relevant contribution. The temperature-dependent part of the phase fluctuations is free of ultra-violet divergences and given by
\begin{align}
\langle[&\theta({\bf x},0) - \theta({\bf 0},0)]^{2} \rangle = -\frac{2 T^{\mathrm{2B}}}{\pi} \int_{0}^{\infty} dk\, k \\ \nonumber
& \int_{0}^{\infty} d(\hbar \omega) \rho_{\mathrm{B}}(k, \omega)  N_{{\mathrm{BE}}}(\hbar \omega)  \left\{1 -   J_{0}(k x) \right\}.
\end{align}
We evaluate this quantity for $\beta n_{0} T^{\mathrm{2B}} \simeq 1.3 \cdot 10^{-2}$. This corresponds to a typical value for BEC of photons in the regime where the density fluctuations are suppressed and we can focus on the phase fluctuations \cite{Klaers}.  
\begin{figure}[t]
 \centerline{\includegraphics[scale=1.14]{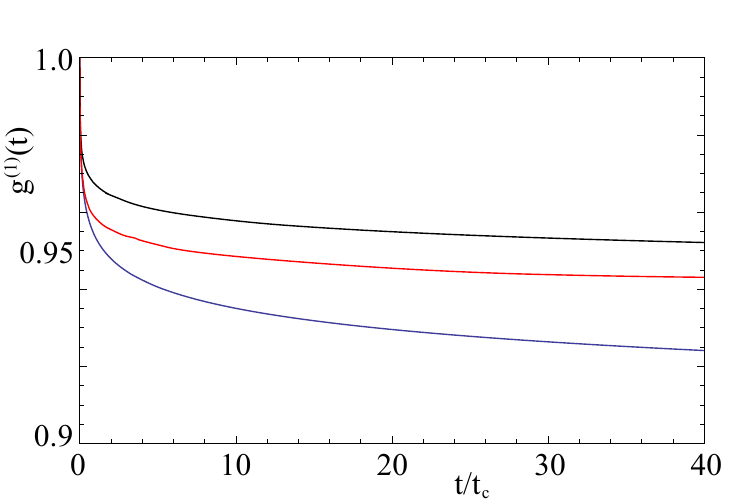}}
 \caption{The normalized first-order correlation function $g^{(1)}(t)$ for a two-dimensional Bose-Einstein condensate as a function of $t / t_{\mathrm{c}}$, with $t_{\mathrm{c}} = \hbar   (n_{0} T^{\mathrm{2B}} \sqrt{\beta})^{-2} \simeq 1.5 \cdot 10^{-10} \, \mathrm{s}$. Here $\beta n_{0} T^{\mathrm{2B}} \simeq 1.3 \cdot 10^{-2}$ and the black, red and blue curve are for respectively $\alpha = 10^{-1}$, $\alpha = 10^{-2}$ and $\alpha = 0$.} 
 \label{fig:TimeCor2}
\end{figure}
By looking at Fig.\,\ref{fig:PhaseCorTemp}, we observe that the temperature-dependent contribution is indeed several orders of magnitude larger than the zero-temperature part. Furthermore, it turns out that the $\alpha$ dependence of the non-zero temperature part is negligible. 
\newline
\indent In order to understand this feature, we distinguish between two different frequency regimes. Namely, $\beta \hbar \omega < 1$ and  $\beta \hbar \omega > 1$. Since we are at room temperature, the latter corresponds to relatively high values of the momentum $k$. In the Bose-Einstein condensed phase the contributions for small momenta are dominant. Therefore, the contributions coming from $\beta \hbar \omega > 1$ are suppressed, and we can focus on the first regime. 
\newline
\indent To make analytical progress, we use that $2 N_{{\mathrm{BE}}}(\hbar \omega) \simeq 2 / \beta \hbar \omega - 1$ for $\beta \hbar \omega < 1$. Furthermore, we can neglect the $-1$ since this is a contribution of the same order as the zero-temperature part and is therefore neglible compared the temperature-dependent part. Hence, we obtain for the non-zero temperature part of the phase fluctuations
\begin{align}
\langle[&\theta({\bf x},0) - \theta({\bf 0},0)]^{2} \rangle = -\frac{2 T^{\mathrm{2B}}}{\pi \beta} \int_{0}^{\infty} dk\, k \\ \nonumber
&\int_{0}^{\infty} d(\hbar \omega) \frac{\rho_{\mathrm{B}}(k, \omega) }{\hbar\omega} \left\{1 -   J_{0}(k x) \right\}.
\end{align}
We can perform the integral over $\omega$ analytically, and we obtain
\begin{align}\label{eq:phasetemp}
\langle[&\theta({\bf x},0) - \theta({\bf 0},0)]^{2} \rangle = \frac{T^{\mathrm{2B}}}{\pi \beta} \int_{0}^{\infty} dk\, k \frac{1 - J_{0}(k x)}{(\hbar \omega_{\mathrm{B}}(k))^{2}}.
\end{align}
This expression is indeed independent of $\alpha$, and this explains why the $\alpha$ dependence of the spatial phase fluctuations is negligible. Note that this argument is independent of the number of dimensions, and therefore also in one dimension the $\alpha$ dependence of the spatial correlations is negligible. 
\begin{figure}[t]
 \centerline{\includegraphics[scale=1.14]{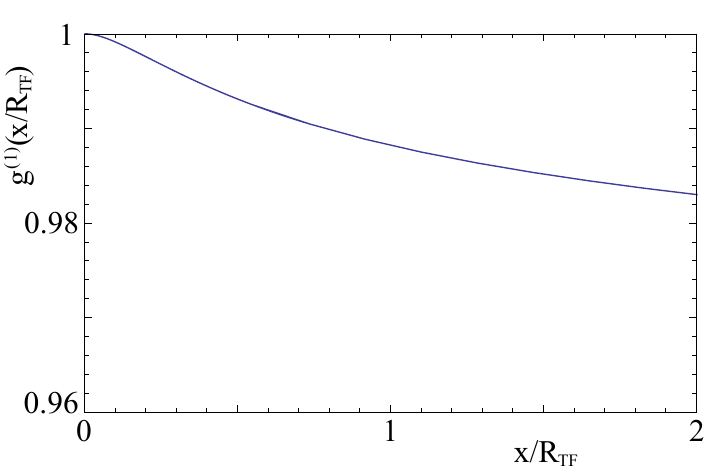}}
 \caption{The normalized first-order correlation function $g^{(1)}(x/R_{\mathrm{TF}})$ for the Bose-Einstein condensate of photons for a condensate fraction $N_{0} / N = 0.2$.} 
 \label{fig:TrueCon}
\end{figure}

\subsection{Temporal correlations}
In order to study the temporal correlation function, we define the first-order correlation function $g^{(1)}({\bf x},t)$ as follows
\begin{align}\label{eq:corg}
g^{(1)}({\bf x},t) := \frac{\langle \phi^{*}({\bf x},t) \phi({\bf 0}, 0) \rangle}{\langle |\phi({\bf 0},0)|^{2} \rangle},
\end{align}
with the temporal correlations defined as $g^{(1)}(t) = g^{(1)}({\bf 0},t)$. Similarly to the spatial correlations, we consider the regime in which the phase fluctuations are dominant. Therefore, we can directly calculate $g^{(1)}(t)$ by using Eqs.\,\eqref{eq:phasegone} and \eqref{eq:phasegamma2}, where we again consider $\beta n_{0} T^{\mathrm{2B}} \simeq 1.3 \cdot 10^{-2}$. In Fig.\,\ref{fig:TimeCor2} we show $g^{(1)}(t)$ in two dimensions for several values of $\alpha$. As can be seen from the figure, $g^{(1)}(t)$ increases for increasing $\alpha$. Furthermore, we find the same qualitative behaviour in one dimension. Thus as opposed to the spatial correlations, the finite lifetime effects are important for the temporal correlations. 

\section{Photons}
\label{sec:normal}
In the previous section we gave a general discussion on finite lifetime effects on correlation functions of bosons in the Bose-Einstein condensed phase. In this section we will focus on a specific example of such a system, namely BEC of photons \cite{BECphoton}. Since this system is two-dimensional, we first investigate whether the photons form a quasi-condensate or a true condensate. 

\subsection{One-particle density matrix}
In order to determine whether the photons form a quasi-condensate or a true condensate, we need to calculate the off-diagonal long-range behaviour of the one-particle density matrix, and compare the size of the condensate with the distance over which the one-particle density matrix falls off. In particular, if the size of the condensate is smaller than the distance over which the one-particle density matrix reduces to say half of the maximum value, we have a true condensate. Otherwise, we are in the quasi-condensate regime. Furthermore, we consider the regime where the density fluctuations are surpressed and therefore we are at large condensate fractions. Therefore we are allowed to use the Thomas-Fermi approximation, and we obtain for the number of condensed photons
\begin{align}\label{eq:TF}
N_{0} = \frac{2 \pi}{T^{\mathrm{2B}}} \int_{0}^{R_{\mathrm{TF}}} dr \, r \left(\mu - \frac{1}{2} m \Omega^{2} r^{2} \right),
\end{align}
where the Thomas-Fermi radius $R_{\mathrm{TF}} = \sqrt{2 \mu / m \Omega^{2}}$ is the size of the condensate. Note that the constant energy $m c^{2}$, with $c$ the speed of the photons in the medium, is absorbed in the definition of the chemical potential $\mu$. Furthermore, $N_{0}$ can be related to the total number of photons $N$ according to \cite{PS}
\begin{align}\label{eq:CF}
N_{0} = N - \frac{\pi^{2}}{3} \left(\frac{k_{\mathrm{B}} T}{\hbar \Omega} \right)^{2}.
\end{align}
By performing the integral in Eq.\,\eqref{eq:TF}, we can relate the chemical potential to the total number of photons in our system. 
\begin{figure}[t]
 \centerline{\includegraphics[scale=1.14]{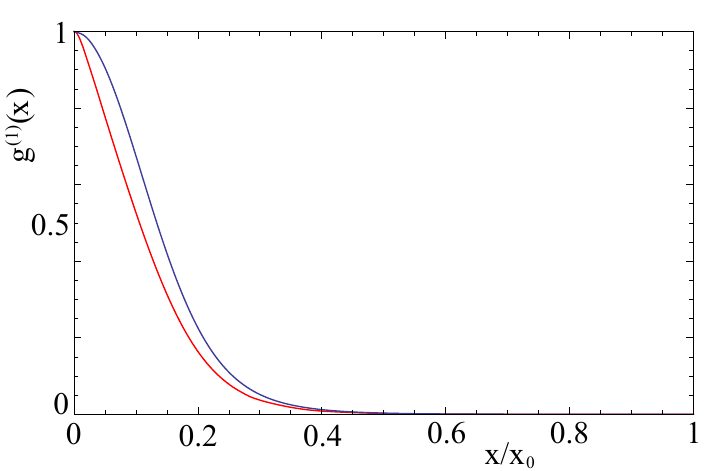}}
 \caption{The normalized first-order correlation function $g^{(1)}(x)$ of the photons in the normal state for $\mu = 0.99 \mu_{\mathrm{c}}$ as a function of $x/x_{0}$, where $x_{0} = \hbar \beta c \simeq 6 \cdot 10^{-6} \,\mathrm{m}$. The blue curve is the result without taking into account the molecules and the red curve corresponds to a molecular density $n_{\mathrm{m}} = 4.5 \cdot 10^{24}\,\mathrm{m}^{-3}$.} 
 \label{fig:NSSpatial}
\end{figure}
This then implies
\begin{align}\label{eq:RTF}
R_{\mathrm{TF}} = \left(\frac{4 T^{\mathrm{2B}} N_{0}}{\pi m \Omega^{2}} \right)^{1/4}.
\end{align}
Furthermore, for the density of condensed photons $n_{0}$ we take the density in the center of the trap. Hence,
\begin{align}\label{eq:DTF}
n_{0} = \sqrt{\frac{m \Omega^{2} N_{0}}{\pi T^{\mathrm{2B}}}}.
\end{align}
Experimentally, the relevant parameter is the condensate fraction $N_{0} / N$. 
\begin{figure*}[t]
  \centerline{\includegraphics[scale=1.14]{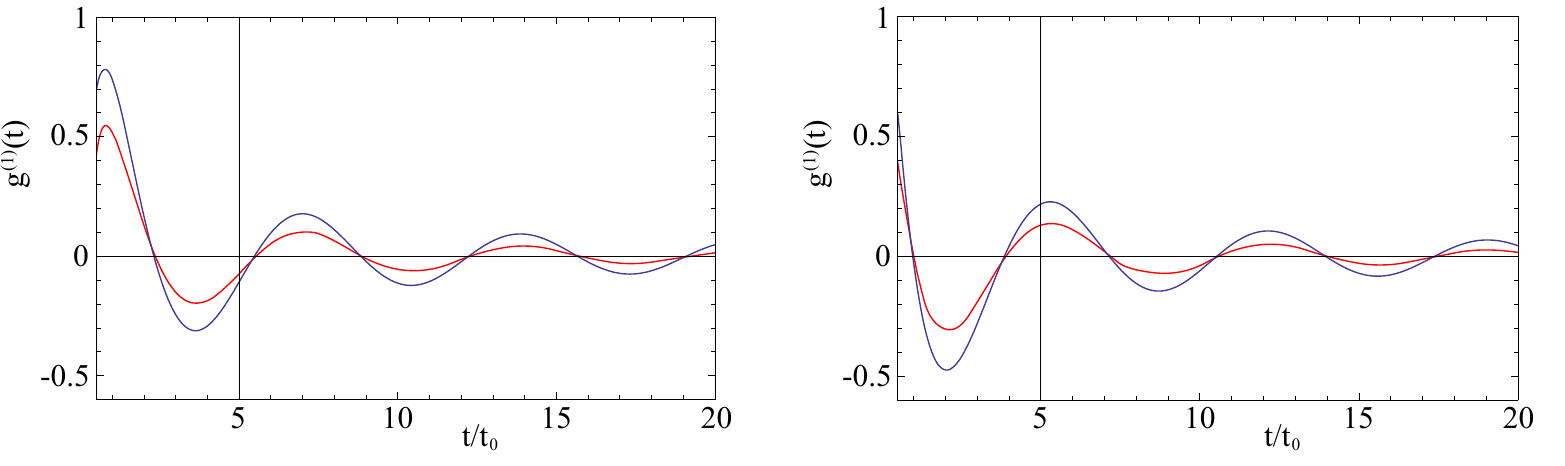}}
\vspace{-0.1 cm}
    \caption{The real (left) and the imaginary (right) part of the normalized first-order temporal correlation function $g^{(1)}(t)$ for the photons in the normal state for $\mu = 0.99 \mu_{\mathrm{c}}$ as a function of $t / t_{0}$ with $t_{0} = \beta \hbar \simeq 2.5 \cdot 10^{-14} \, \mathrm{s}$. For the blue curve we omitted the effect of the molecules and the red curve corresponds to the molecular density $n_{\mathrm{m}} = 4.5 \cdot 10^{24}\, \mathrm{m}^{-3}$.}
     \label{fig:NSTemp}
\end{figure*}
Therefore given a condensate fraction, we use Eq.\,\eqref{eq:CF} to determine $N_{0}$ and with this value we obtain the size and the density of the condensate via Eqs.\,\eqref{eq:RTF} and \eqref{eq:DTF}. Furthermore, we use Ref.\,\cite{BECphoton} to obtain relevant numerical values for the parameters $m$, $\Omega$ and $T^{\mathrm{2B}}$.
\newline
\indent In Fig.\,\ref{fig:TrueCon} we show a plot of the normalized first-order spatial correlation function $g^{(1)}(x/R_{\mathrm{TF}}) = g^{(1)}(x/R_{\mathrm{TF}},0)$ as defined in Eq.\,\eqref{eq:corg} for a condensate fraction of $20\,\%$. From this plot it is clear that the phase correlation function $g^{(1)}({\bf x})$ hardly drops over the condensate size, and that the photons form a true condensate. In principal we should also include the harmonic trap in the calculation for $g^{(1)}(x/R_{\mathrm{TF}})$, but we are so far in the true condensate regime that this correction will not influence this conclusion.

\subsection{Normal state}
Apart from the Bose-Einstein condensed phase, the photons can also be in the normal state. Since in this case not only the small frequency behaviour of the self-energy is important, the details of the system of interest should be included and we need the exact expression for the self-energy. Therefore, we use the explicit expression for the self-energy as given in Ref.\,\cite{AW}. In the normal state we write for the first-order correlation function  
\begin{align}
\langle \phi^{*}({\bf x},\tau) \phi({\bf 0},0) \rangle = \frac{1}{\hbar\beta V} \sum_{{\bf k},n} G({\bf k},i \omega_{n}) e^{i (\omega_{n} \tau - {\bf k} \cdot {\bf x})},
\end{align}
where
\begin{align}
G({\bf k},i \omega_{n}) = \frac{-\hbar}{-i \hbar \omega_{n} + \epsilon({\bf k}) - \mu + \hbar\Sigma({\bf k},i \omega_{n})}.
\end{align}
Note that in this calculation we neglect the self-interaction of the photons, since we are primarily interested in the effect of the imaginary part of the self-energy. By defining
\begin{align}
\rho({\bf k},\omega) := \frac{1}{\pi \hbar} \mathrm{Im}\left[G({\bf k},\omega^{+}) \right],
\end{align}
we write
\begin{align}
\langle &\phi^{*}({\bf x},t) \phi({\bf 0},0) \rangle \\ \nonumber
&= \frac{1}{2 \pi} \int dk \int d(\hbar\omega) \, k\,\rho({\bf k},\omega) N_{\mathrm{BE}}(\hbar\omega) J_{0}(k x) e^{i \omega t} ,
\end{align}
where $J_{0}(k x)$ is the Bessel function of the first kind, $N_{\mathrm{BE}}(\hbar\omega)$ is the Bose-Einstein distribution function as defined in Eq.\,\eqref{eq:BED} and $x = |{\bf x}|$. Now we study the spatial and temporal correlation functions $g^{(1)}(t)$ and $g^{(1)}(x)$ seperately. In general we are interested in the regime where we are close to condensation, and therefore we take $\mu \simeq 0.99 \mu_{\mathrm{c}}$. Furthermore, we take the parameters as in the experiment of Ref.\,\cite{BECphoton}. It turns out that for the densities used in these experiments the effect of the molecules is small. In order to demonstrate the effect of the dye molecules we take $n_{\mathrm{m}} = 4.5 \cdot 10^{24}\, \mathrm{m}^{-3}$. In general high molecular densities can spoil the thermalization of the photons, but this value should be within the regime in which the photons can still thermalize.
\newline
\indent As can be seen in Fig.\,\ref{fig:NSSpatial}, the normalized spatial correlation function $g^{(1)}(x)$ is lowered by the effect of the molecules. In order to make a prediction for the experiment we should incorporate the harmonic potential. However, the harmonic oscillator length of the trap is roughly $8 \, \mu\mathrm{m}$ and thus sufficiently larger than the distances over which $g^{(1)}(x)$ is non-zero. Therefore, we expect that the incorporation of this harmonic potential is a small effect. 
\newline
\indent The normalized first-order temporal correlation $g^{(1)}(t)$ consist of an imaginary and a real part, which we show seperately in Fig.\,\ref{fig:NSTemp}. For both parts the amplitude of the oscillations are decreased by the interaction with the molecules. Here, we used correlation functions in terms of creation and annihilation operators. In experiments one measures the correlation between the electric field at different times and positions, and therefore experimentally only the real part is relevant.

\section{Conclusion and outlook}
\label{sec:concl}
In this paper we investigated finite lifetime effects, characterized by the dimensionless parameter $\alpha$, on the first-order correlation functions. By taking into account the phase fluctuations up to all orders, we derived an explicit expression for the first-order correlation functions in the Bose-Einstein condensed phase for high condensate fractions. We showed that the value of $\alpha$ does not influence the spatial correlations, but it enhances the temporal correlation function. 
\newline
\indent Subsequently, we focussed on BEC of photons under the relevant experimental conditions and we showed that the phase of the condensate is coherent over length scales larger than the size of the condensate. Therefore, the photons form a true condensate. Finally, we calculated the normalized first-order correlation functions of the photons in the normal state and we showed that the spatial and temporal correlations are both surpressed by the interaction with the dye molecules. 
\newline
\indent For future research it is interesting to investigate the regime with smaller condensate fractions. Here the density fluctuations are important and they also have to be incorporated in the formalism. For BEC of photons this regime is also accessible experimentally \cite{Klaers}, and in this case the effect of the interaction with the dye molecules can be different from the case with high condensate fractions.

\acknowledgments
It is a pleasure to thank Jan Klaers and Martin Weitz for useful discussions. This work was supported by the Stichting voor Fundamenteel Onderzoek der Materie (FOM), the European Research Council (ERC) and is part of the D-ITP consortium, a program of the Netherlands Organisation for Scientific Research (NWO) that is funded by the Dutch Ministry of Education, Culture and Science (OCW)

\end{document}